# Observational Diagnostics for Two-Fluid Turbulence in Molecular Clouds as Suggested by Simulations

By

Chad D. Meyer[1], Dinshaw S. Balsara[1], Blakesley Burkhart[2] and Alex Lazarian[2]

[1]Physics Department, University of Notre Dame

[2]Astronomy Department, University of Wisconsin at Madison

**Abstract**

We present high resolution simulations of two-fluid (ion-neutral) magnetohydrodynamic (MHD) turbulence with resolutions as large as $512^3$. The simulations are supersonic and mildly sub-Alfvénic, in keeping with the conditions that arise in molecular clouds. Such turbulence is thought to influence star formation processes in molecular clouds because typical cores form on length scales that are comparable to the dissipation scales of this turbulence in the ions. The simulations are motivated by the fact that recent studies of isophotologue lines in molecular clouds (Li & Houde 2008, Hezareh *et al.* 2010) have found significant differences in the linewidth-size relationship for neutral and ion species. Our first goal in this paper is to explain those observations using simulations and analytic theory (Balsara 1996). Our second goal in this paper is to present a new set of density-based diagnostics by drawing on similar diagnostics that have been obtained by studying single-fluid turbulence (Burkhart *et al*. 2009, 2010). We further show that our two-fluid simulations play a vital role in reconciling alternative models of star formation.

The velocity-dependent diagnostics display a very interesting complementarity with the density-dependent diagnostics. We find that the linewidth-size relationship should show a prominent difference between ions and neutrals when the line of sight is orthogonal to the mean field. This is because the MHD waves in the ions differ strongly from hydrodynamic waves in the neutrals on the dissipation scales of the two-fluid turbulence. Such MHD waves have a strongly preferred velocity fluctuation that is orthogonal to the mean field. We also find that the density probability distribution functions (PDFs) and their derived diagnostics should show prominent differences between the ions and neutrals when the line of sight is parallel to the mean field. This is because the preferentially transverse velocity fluctuations in the ions (at dissipation scale) produce density fluctuations that are most prominently visible when viewed along the field lines. When the magnetic field makes an angle to the line of sight, both linewidth-size differences and density PDF differences should be visible in the ions and neutrals. The diagnostics presented here should be easy for observers to test. The present analysis is predicated on observations in optically thin lines and assumes a mean magnetic field that does not change direction within a cloud.

**1) Introduction**

Molecular clouds are the densest component of the multiphase interstellar medium (ISM). Star formation occurs almost solely in molecular clouds. In fact, almost all molecular clouds seem to show evidence of star formation (Blitz 1993). Understanding the host environment in which stars form is important in understanding the processes that influence star formation.

Molecular clouds are known to be strongly turbulent environments. The first evidence came from the scaling relations of Larson (1981). A recent reexamination of the scaling relations by Heyer & Brunt (2004) shows that this turbulence is universal to all molecular clouds, regardless of their current star formation rate or surrounding galactic environment. This suggests that the source of the turbulence could be large scale driving by energetic processes which are common to all molecular clouds. Important information was obtained by studying the velocity power spectrum of turbulence in diffuse media using Velocity Channel Analysis (VCA) and Velocity Correlation Spectrum (VCS) techniques (Lazarian & Pogosyan 2000, 2004, 2006), which revealed steep spectra of velocity turbulence in diffuse gas (Chepurnov et al. 2010) and molecular gas (Padoan et al. 2009). A compilation of the results which have been obtained by these techniques is reviewed in Lazarian (2010). The observed velocity dispersion increases with length scale and is supersonic for all but the shortest observed lengths ($L \gtrsim 0.05$ pc) (Heyer & Brunt 2004). On large scales, the turbulence can have an RMS mach number of 5 or greater (Crutcher 1999, Kainulanen & Tan 2013). This results in highly compressible turbulence. The spatial distribution of Mach numbers for the SMC is obtained in Burkhart et al. 2010 using HI data. With synchrotron emission, one obtains evidence for transonic turbulence (Gaensler et al. 2011, Burkhart, Lazarian & Gaensler 2012), which agrees well with independent Hα studies in Hill et al. (2008).

Molecular clouds and the cores that form in them are also strongly magnetized. Zeeman observations indicate large magnetic field components along the line of sight of the observer. Initial studies of Zeeman measurements by Crutcher (1999) indicated that cores which have formed in molecular clouds are consistent with being magnetically critical or super-critical, meaning that the magnetic field is not strong enough to prevent gravitational collapse. More recent results (Crutcher et al. 2010) confirm that observed clouds and cores have small magnetic fields and are consistently super-critical over many observations. They find that the distribution of maximal values of line-of-sight magnetic field scales as $n^{2/3}$, which excludes sub-critical cores with magnetic support.

Molecular clouds also show a very low ionization fraction. This is important because only electrons and ions couple to the magnetic fields. The outer regions of clouds are not shielded from ionizing UV radiation, but the denser inner regions of the clouds are ionized by cosmic rays. McKee (1989) discussed the different potential components of molecular clouds and their corresponding recombination rates, and derived a general expression for the ionization fraction for cosmic ray ionized clouds, $\chi(e) = 1.3 \times 10^{-5} n(\mathrm{H}_2)^{-0.5}$. Caselli et al. (2002) showed that the freeze-out of molecules on dust grains can significantly decrease the ionization fraction at higher densities. McKee (1989) approximates the effects of both far UV radiation and cosmic rays, and shows that the far UV radiation will dominate the ionization rate at low densities. However, even for the lowest density regions of molecular clouds, with n~100/cm$^3$, the ionization fraction is predicted to be about $6 \times 10^{-7}$.

There are three leading models for the early stages of star formation. We mention them here in the chronological order in which they were developed. The first model is the magnetically regulated model proposed in Mestel & Spitzer (1956), Mouschovias (1977), Shu (1983), Ciolek & Mouschovias (1994), Basu et al. (2009), Dapp et al. (2012). In this model, cloud material collapses gravitationally into clumps and cores which are magnetically supported against collapse. The magnetic flux slowly leaks out of the core due to ambipolar diffusion, until the core becomes super-critical to collapse. A major criticism of this model is the lack of observational evidence for magnetically sub-critical cores (Crutcher, 1999; Bourke et al., 2001; Troland & Crutcher, 2008; Crutcher et al., 2010), although this may not be conclusive (for instance Mouschovias & Tassis, 2009; 2010). In the second model, the turbulence mediated star formation model, clumps and cores form at the intersection of colliding flows due to the over-densities produced by supersonic, compressible turbulence. Several ideal MHD-based turbulence simulations have been presented in the literature starting from (Klessen, Heitsch & Mac Low 2000, Ostriker, Stone & Gammie 2001, Balsara, Ward-Thompson & Crutcher 2001) and going on most recently to Collins et al. (2012). Ideal magnetohydrodynamic (MHD) simulations have shown that the properties of cores can be reproduced via turbulent simulations, though it requires small magnetic fields (Tilley & Pudritz, 2004, 2007). The third model is based on the realization that turbulence can enhance magnetic reconnection rates (Lazarian & Vishniac, 1999) and is, therefore, known as the reconnection diffusion model (Lazarian 2011, Santos-Lima et al. 2010, 2013). In this model, the turbulence-enhanced reconnection allows the magnetic flux to diffuse out of cores, thereby facilitating core collapse and subsequent star formation (Lazarian, Esquivel & Crutcher 2012). The fast reconnection theory also provides more justification to the second approach. Indeed, the reconnection in computer simulations is generically fast due to numerical resistivity, which poses an important question as to what extent the numerical simulation results can be trusted in terms of the magnetic field dynamics. The theory in Lazarian & Vishniac (1999) demonstrates that magnetic reconnection in turbulent media happens at the eddy turnover time, independent of resistivity and therefore the enormous differences of the magnetic Reynolds number, or Llundquist number $S = LV_A/\nu$ (where $V_A$ is the Alfven velocity and $\nu$ is the Ohmic diffusivity) do not change the dynamics of magnetized eddies. In fact, the Lazarian & Vishniac (1999) theory proves that the magnetic field is not frozen in a turbulent medium. This conclusion was rigorously proven in Eyink, Lazarian & Vishniac (2011), where the reconnection theory was connected with the concept of Richardson diffusion. The testing of the Lazarian & Vishniac (1999) theory can be found in Kowal et al. (2009, 2012), while the numerical confirmation of the flux being not frozen is provided in Eyink et al. (2013).

Numerical simulations have shown themselves to be useful in understanding the nature of compressible MHD turbulence. In particular, they can bridge the gap between theory and observation. The spectral index of the turbulence is related to the structure function of the velocity (Elmergreen & Scalo, 2004), and simulations report many different values of spectral index (see the review by Ballesteros-Paredes et al., 2007, Burkhart *et al.* 2010). Results vary based on Mach number and approach *n=-2* for high Mach number flows (Cho & Lazarian, 2003; Ballesteros-Paredes et al., 2006), which is the spectrum for shock-dominated flow (Elsasser & Schamel, 1976). Burkhart *et al.* (2010) shows the relationship between spectral index and sonic Mach number for different magnetic field values. Density studies are also useful in differentiating flow features. Kowal, Lazarian & Beresnyak (2007) show that the column density probability distribution function (PDF) varies with sonic Mach number and could be used as an observational diagnostic. Burkhart *et al.* (2009) confirms these results with higher resolution simulations and extends the analysis to include varying mean magnetic field values.

Recent observations of molecular clouds have revealed noticeable line width differences between ionized and neutral molecules (Li & Houde 2008, Li et al. 2010). The line widths of the ionized molecules were found to be narrower than the neutral molecules along many lines of sight, and in particular the narrowest ion lines were narrower than the narrowest neutral lines along every length scale they observed. The observations by Li et al. (2010) are especially valuable because they were carried out using isophotologue molecules with the $^{13}$C isotope, specifically the 4-3 transition of HCO+ and HCN. This enabled them to probe the full extent of the molecular cloud. By now, this effect has been observed in several clouds, including M17, DR 21(OH), NGC 2024, Cygnus X (Li et al. 2010; Hezareh et al.

2010). By drawing on an analysis of wave propagation in molecular clouds by Balsara (1996), Li & Houde (2008) realized that two-fluid effects strongly influence the evolution of the turbulence. In particular, Balsara (1996) had shown that in any two fluid plasma there exists a length scale, $L_{AD} = v_A / \alpha \rho_i$, below which the propagation of Alfvén waves and specific families of magnetosonic waves is strongly damped. (Here $L_{AD}$ is the threshold length scale for ambipolar diffusion, $v_A$ is the Alfvén speed, $\alpha$ is the frictional coupling coefficient between the ions and neutrals and $\rho_i$ is the density of ions. We adopt the physical value of $\alpha = 9.2 \times 10^{13} \text{cm}^3 \text{s}^{-1} \text{g}^{-1}$ from Smith & Mac Low (1997).) For fiducial molecular cloud parameters with number density $10^3$ atoms/cm$^3$, magnetic field of 15 $\mu$G and ionization fraction of $10^{-7}$, we get $L_{AD} = 0.05 \text{ pc}$. Since the Alfvén waves and magnetosonic waves are strongly damped on length scales smaller than $L_{AD}$, the turbulence in the ions is extinguished on these smaller scales. The turbulence in the neutrals continues unabated, with the result that the neutral line profiles are broader than the line profiles in the ions. The dispersion analysis in Balsara (1996) was restricted to isothermal two-fluid MHD, but it has since been extended to two-fluid MHD in molecular clouds with realistic heating and cooling (Tilley & Balsara 2011). The more detailed analysis from Tilley & Balsara (2011) continues to support the interpretation of linewidth-size relationship that was provided by Li & Houde (2008).

Simulating these phenomena requires an authentic two-fluid approach. Single fluid MHD with an ambipolar diffusion operator applied to the magnetic fields (O'Sullivan & Downes, 2006; 2007) does not retain the different momentum and density distributions of the ions and neutrals, and so would not be able to demonstrate the line width differences observed by Li & Houde (2008) and Li et al. (2010). The full two-fluid approach is computationally very challenging. The maximum stable time-step is limited by the fastest wave speed in the problem, which can be approximated by the Alfvén speed in the ions. This speed is proportional to $\rho_i^{-1/2}$, thus very short time-steps must be taken when the ion density is very small, as in molecular clouds. An approach that is often employed to lower the computational cost of these simulations is the heavy ion approximation (HIA) (Oishi & Mac Low 2006; Li et al. 2008). In this case the molecular mass of the ions is artificially increased to decrease the Alfvén speed, while at the same time decreasing the coupling constant so that the friction term is not changed. Tilley & Balsara (2010) showed in the dispersion analysis of HIA that unphysical waves will emerge in the ions on length scales shorter than the ambipolar diffusion scale. This is undesirable in turbulence simulations, because shorter wavelength perturbations will not be adequately damped by the drag between the ions and neutrals, which would change the turbulent spectrum on these length scales. Therefore, we find the full two-fluid approach tracing realistic ion masses to be the most useful in simulating the magnetized turbulence in molecular clouds.

Recent simulations (Tilley & Balsara, 2010) have sought to examine the nature of two-fluid turbulence without the HIA and reproduce the observations of Li & Houde (2008) and Li et al. (2010). The results of these simulations have shown that this model is able to reproduce the observation. Specifically, Tilley & Balsara, (2010) showed that synthetic line widths produced using their simulations yielded consistently narrower line widths in the ions than in the neutrals. However, these simulations were limited in resolution to $192^3$ zones.

The goal of this paper is to extend the simulations in Tilley & Balsara (2010) to substantially higher resolution and to incorporate an extended range of magnetic fields. We show that the results persist in the higher resolution simulations presented here. Our second goal is to extend some of the statistical methods presented in Burkhart et al. (2011) to two-fluid, partially ionized turbulence. Our third and most important goal is to use the simulations to extract several diagnostics that can be used quite easily by observers to diagnose the existence of two-fluid effects in turbulent molecular clouds. Some of the diagnostic tools are new and have not yet been applied to observed data sets. Such an application would indeed be very valuable for studies of star forming molecular clouds.

The remainder of this paper is organized into the following sections. In Section 2, we discuss the numerical methods employed in these simulations. In Section 3, we present the results of spectral analysis and synthetic line widths. In other words, Section 3 presents diagnostics that are determined for the most part by the turbulent velocities. In Section 4 we present bulk results including density PDFs and the skewness and kurtosis of the density maps. Section 4, therefore, focuses on diagnostics that are mostly determined by the densities in the turbulent flow. In Section 5, we discuss the results and present our conclusions.

**2) Methods**

We use the RIEMANN code (Balsara 1998a,b, 2004; Balsara & Spicer 1999a,b) to update the neutral and ionized fluid variables. The neutral fluid obeys the isothermal Euler equations, the ionized fluid obeys the isothermal MHD equations, and the ion-neutral drag term is incorporated using an implicit, operator-split method (Tilley & Balsara, 2008, Tilley, Balsara & Meyer 2011). The numerical method is very similar to Tilley & Balsara (2010). In this application, we have chosen to use a centered

WENO reconstruction and linearized Riemann solver in order to reduce the numerical dissipation present in the simulation. The WENO methods are based on Jiang & Shu (1996) and Balsara & Shu (2000).

The fluids are initially uniform in density, with the neutral fluid set to unity. The mean molecular weight is set to $\mu_n = 2.3$ amu for the neutral fluid and $\mu_i = 29$ amu for the ionized fluid. Thus, for a particular ionization fraction $\chi$, the ion density will be given by $\rho_i = \chi \rho_n \mu_i / \mu_n$. The mean magnetic field is initialized to be uniform in the $x$-direction, with the other components initially zero. The velocity field is initially the Fourier transform of a random-phased Gaussian peaked at $k = 2$ and spanning $1 \leq k \leq 4$, with an RMS velocity of 2.5 times the sound speed. The forcing function has the same structure as the initial velocity, but with a different random phase. The phases used change episodically as time progresses, ensuring that the turbulent forcing will not leave an imprint on the simulation. The forces are applied to both the ions and the neutrals with the same acceleration to maintain a constant kinetic energy throughout the simulation.

In this paper, we present two forced turbulence simulations with strong magnetic fields on $512^3$ zone meshes. We set the Alfvén speed for the combined fluid to be 3 and 6 times the sound speed; we refer to these runs as A3 and A6 respectively. Finally, we set the coupling coefficient between the ions and neutrals. In our simulations, we scale the parameters such that the ambipolar diffusion length scale, $L_{AD} = v_A / \alpha \rho_i$, is 80 zones. In this way, we have a clear separation between the driving scale, which is at long wavelengths (512-128 zones), the ambipolar diffusion scale (80 zones), and the numerical dissipation scale, which is at short wavelengths (about 20 zones in our code). This permits us to accurately model the turbulence both above and below the ambipolar diffusion length scale without contamination from numerical effects. In order to have the same $L_{AD}$ for both simulations, we use the same value of coupling constant, $\alpha$, and different ionization fractions for each. For the simulation with Alfvén speed of 3, we have an ionization fraction of $10^{-4}$, and for the simulation with Alfvén speed of 6, the corresponding ionization fraction is $2 \times 10^{-4}$. By matching $L_{AD}$ across the two simulations, this pair of simulations is well suited to pick out differences in the turbulence that result from the differing magnetic field strength.

We also include a third forced turbulence simulation with a smaller magnetic field such that the Alfvén speed is equal to the sound speed. We call this run A1 and it was also done on a $512^3$ zone mesh. While cloud cores are predominantly sub-Alfvénic environments, the envelope regions have been shown to have Alfvénic Mach numbers greater than unity (see Lazarian, Esquivel & Crutcher 2012). Furthermore, the overall Alfvénic nature of giant molecular clouds on the largest scales of the cloud is still strongly debated, but a wide range of evidence constructs a picture of globally super-Alfvénic clouds (Padoan & Nordlund 1999; Lunttila et al. 2008; Crutcher et al. 2009; Burkhart et al. 2013) Additionally, run A1 serves as a comparison between the low and high magnetic field regimes. This simulation has an ionization fraction of $\chi = 10^{-4}$. The driving is as above, but in this case the RMS velocity of the turbulence is 3 times the sound speed. The Alfvén speed is equal to the sound speed. The dissipation scale is set to 40 zones. All of the simulation data is summarized in Table 1.

**Table 1** – Summary of parameters for the $512^3$ zone simulations presented in this paper.

| Run Name | Alfvén Speed | Sonic Mach # | Alfvénic Mach # | Ionization Fraction | Ambipolar Diffusion Scale | Driving Scale |
| --- | --- | --- | --- | --- | --- | --- |
| A3 | 3.0 | 2.5 | 0.83 | $10^{-4}$ | 80 Δx | 128 Δx |
| A6 | 6.0 | 2.5 | 0.42 | $2\times10^{-4}$ | 80 Δx | 128 Δx |
| A1 | 1.0 | 3.0 | 3.0 | $10^{-4}$ | 40 Δx | 128 Δx |

**3) Line Widths and Spectra**

In this section, we present analysis of the velocity-dependent features of the simulations described in Section 2 above. We will show simulated line widths and linewidth-size relationships, for comparison with observations (Li & Houde 2008, Li et al. 2010, Hezareh et al. 2010). We will also show kinetic and magnetic energy spectra for the simulations.

Sub-Section 3.1 presents the energy spectrum and discusses it in terms of the dissipation properties of the waves. Sub-section 3.2 focuses on the dissipation scales in the turbulence. Sub-Section 3.3 presents simulated line profiles. Sub-Section 3.4 discusses linewidth-size relationships in the two-fluid turbulence.

**3.1) Energy spectrum**

Balsara (1996), and Tilley & Balsara (2008, 2010) have undertaken the dispersion analysis for the isothermal, two-fluid MHD system. This analysis demonstrates that on length scales shorter than the ambipolar diffusion scale, only one family of MHD waves persists, whereas the others are strongly damped. Specifically, Balsara (1996), and Tilley & Balsara (2011) find that when the Alfvén speed is greater that the sound speed, then the Alfvén waves and fast magnetosonic waves become non-propagating on length scales that are smaller than the ambipolar diffusion scale. Conversely, when the Alfvén speed is less than the sound speed, the Alfvén and show magnetosonic waves become non-propagating for lengths shorter than the ambipolar diffusion scale. When the Alfvén waves die out, the MHD turbulence cannot be sustained. Since the Alfvén waves are strongly damped below the ambipolar diffusion scale, the magnetic energy is expected to fall off below that scale as is the kinetic energy of the ions.

Fig. 1 shows the turbulent kinetic energy spectrum from our simulations for the neutral and ionized fluids as well as the magnetic field. Fig. 1a represents run A3, Fig. 1b shows run A6 and Fig. 1c shows run A1. The driving and ambipolar diffusion scales are labeled on the plots. The black line represents the spectrum of the kinetic energy of the neutrals, $\rho_n v_n^2$, the red solid line represents the kinetic energy of the ions, $\rho_i v_i^2$, and the red dashed line represents the energy in the magnetic fields, $B^2$. The data have been normalized so that the spectra match below the driving scale in the inertial range. The spectra are also compensated by a factor of $k^{5/3}$, which is the expected scaling of hydrodynamical turbulence from Kolmogorov (1941) and also the expected scaling for MHD turbulence from Goldreich & Sridhar (1995). The slope of the inertial range is slightly steeper than this, and, in fact, is quite close to $k^{-2}$, which is the value expected for high Mach number MHD turbulence (Cho & Lazarian, 2003; Ballesteros-Paredes et al., 2006). The simulation parameters we have chosen (see section 2) permit us to separate the scales on which the turbulence is driven, the scales on which ambipolar diffusion acts and the scales on which numerical dissipation dominates in our code. This is essential to demonstrate numerically the physical effect ambipolar diffusion would have on a molecular cloud. On scales that are substantially larger than the ambipolar diffusion length scale, the ions and neutrals are strongly coupled. As a result, on very large length scales, the two-fluid plasma mimics single-fluid MHD turbulence. In that limit, the ions and neutral spectra match closely. Dispersion analysis from Balsara (1996) shows that there is a large range of length scales, from around $10L_{AD}$ to $L_{AD}$, where the coupling between the ions and neutrals is only partial, and the Alfvén and fast magnetosonic wave families are significantly damped in the ions. Wave propagation continues unabated in the neutrals. It is because of this partial coupling that observers cans see a persistent linewidth-size relationship even on length scales that are larger than $L_{AD}$. On length scales that are smaller than the ambipolar diffusion scale, the ions and neutrals are mostly decoupled. Consequently, the ionized fluid has a smaller energy on all scales as compared to the neutral fluid. The effect is slightly larger in the run A6 than in A3. This shows that higher mean magnetic field strength will impose additional drag on the neutral fluid at smaller length scales. Likewise, the effect is considerably smaller for the small magnetic field case of run A1. Balsara (1996) showed that when the magnetic field is large (as in runs A3 and A6), the Alfvén and fast magnetosonic waves are strongly damped, leaving only the slow waves, however in cases where the Alfvén speed is comparable to or less than the sound speed (as in run A1) the Alfvén and slow waves are damped and only the fast waves survive.

Tilley & Balsara (2010) also saw the spectra of the ions and neutrals separate at the ambipolar diffusion length scale (see their Fig. 2). Our simulations are at a much higher resolution, and have the ambipolar diffusion scale set between the driving and numerical dissipation scales. Therefore, we can see a clear (albeit small) inertial range in Fig. 1, even beyond the ambipolar diffusion scale labeled on the plot. This is understandable by recalling that the ambipolar diffusion scale on the plot is for the average of the entire box. High mach number turbulence changes the local properties of the flow by, for instance, creating large density variations. These effects allow the compressible turbulence to have an even larger inertial range than would have been first predicted. The results of this section demonstrate the important role magnetic fields and ambipolar diffusion play in the turbulent spectrum. Specifically, strong magnetic fields cause the turbulent motions of the ions to be less vigorous than the neutrals on a range of length scales centered around the ambipolar diffusion scale.

**3.2) Focus on the Dissipation Scales in the Turbulence**

Fig. 1 shows that the turbulence, as measured by the velocity or magnetic spectra of the ions, extends to length scales well below $L_{AD}$. In this section, we understand that fact by initially using a simple example from hydrodynamic turbulence and then extending it to two-fluid MHD. This is a very important issue because core formation takes place in molecular clouds on length scales that are comparable to $L_{AD}$. Consequently, an understanding of this concept gives us perspective on whether the turbulence can permeate protostellar cores. If it can, it will have an important influence on reconnection or other magnetic processes in the core.

Consider a sound wave that has space-time variation given by $e^{i(kx-\omega t)}$. Here "$k$" is the real wave number and "$\omega$" is the complex angular frequency. Let the sound speed be $c_s$ and the viscosity be $\nu$, and let the fluid have an x-directional velocity $v_x$. It can be shown that the sound waves propagate in the x-direction with

$$\omega = k\left[v_x \pm c_s\sqrt{1 - \frac{4}{9}\frac{k^2\nu^2}{c_s^2}} - i\frac{2}{3}k\nu\right]$$

Thus the sound waves stop propagating with respect to the flow on a dissipative length scale that we can approximate by $L \sim \nu / c_s$. If the fluid is turbulent, and if the turbulence is subsonic, then this is also the length scale at which the turbulence will be truncated. However, if the turbulence is strongly supersonic then the turbulence will truncate on a length scale "$L$" where the Reynolds number, Re, of the turbulence is effectively unity. This is because a Reynolds number of unity characterizes a balance between the advective stresses and the viscous stresses. If we let $v_L$ be the RMS velocity in the turbulence at a length scale "$L$", the dissipation scale sets in on length scales where $\text{Re} = \frac{v_L L}{\nu} = 1$. For strongly supersonic turbulence, this length scale "$L$" can be smaller than $\nu / c_s$.

A Reynolds number-like concept for turbulence that is dominated by ambipolar diffusion was formulated in Balsara (1996). Before we motivate that concept, let us consider an Alfvén wave that has space-time variation as $e^{i(kx-\omega t)}$. Let us take $\rho_i$ to be the ion mass density and $v_{Ax}$ to be the Alfvén speed in the x-direction. It can be shown that the Alfvén waves propagate in the x-direction with

$$\omega = k \left[ v_x \pm v_{Ax} \sqrt{1 - \frac{k^2 \, v_{Ax}^2}{4 \, (\alpha \, \rho_i)^2}} - i \frac{k \, v_{Ax}^2}{2 \, \alpha \, \rho_i} \right]$$

Analogous to the sound waves in the previous paragraph, the Alfvén waves too stop propagating with respect to the flow on a length scale of $L_{AD}$. If the turbulence is subsonic, this is also the length scale at which the turbulence will be truncated. However, because the turbulence is strongly supersonic on larger length scales, the turbulence will truncate on a length scale "$L$" where the Reynolds number of the two-fluid turbulence, $\text{Re}_{AD}$, is effectively unity. If we let $v_L$ be the RMS velocity in the two-fluid turbulence at a length scale "$L$", the dissipation scale sets in on length scales where

$$\text{Re} = \frac{v_L L}{v_{Ax}^2 / (\alpha \, \rho_i)} = 1 \; .$$

For strongly supersonic turbulence, this length scale "$L$" can be much smaller than $L_{AD}$. This is what we see in Fig. 1.

Further insights on whether two-fluid turbulence can permeate scales smaller than $L_{AD}$ is obtained by a structure function analysis. A more detailed structure function analysis has been carried out and will be part of a subsequent paper (Burkhart et al. 2013) where several other theoretical diagnostics of two-fluid turbulence will be presented. Since the present paper is more geared toward observational diagnostics, we only mention the most relevant results from that analysis. It reveals that the anisotropies in the structure function are present even on scales which are somewhat shorter than $L_{AD}$. However, on scales that are much smaller than $L_{AD}$, the anisotropy in the structure functions is absent. If the turbulence were purely ideal MHD turbulence, we would expect a strong anisotropy on all scales; if the turbulence were purely hydrodynamical, we would expect the structure functions to be isotropic on all scales. In our two-fluid simulations, the structure functions change character going as they do from anisotropic on larger scales to isotropic on smaller scales. This shows that there is a range of length scales at and below $L_{AD}$ where the turbulence is changing character, consistent with ambipolar diffusion operating on those length scales.

Now that we have developed this perspective, we realize that the turbulence can permeate to length scales on which cores form. On those length scales, the turbulence can enhance reconnection. Furthermore, since ambipolar diffusion is a dissipative effect on those length scales, it could even enhance the efficiency of ambipolar diffusion, thereby enhancing the efficiency of star formation. This might explain the results of Ward-Thompson, Motte & Andre (1999) who find that star formation in molecular clouds takes place on timescales that are in fact shorter than the classical ambipolar diffusion timescales.

**3.3) Simulated Line Profiles**

The shape and width of spectral line profiles is useful for probing the kinematic structure within a molecular cloud because the line profiles directly track the velocities within the fluid. In Fig. 2, we show synthetic line profiles for our runs taken over the entire domain. The ion lines are shown in red and the neutrals in black. Our line of sight extends all the way through our computational domain. Physically, this amounts to saying that we are observing the cloud in the optically thin lines used by Li & Houde (2008) and Li et al. (2010). Figs. 2(a) – (c) show the profile when observed along the direction of the mean magnetic field for runs A3, A6 and A1 respectively. Figs. 2(d) – (f) show the same when looking perpendicular to the mean magnetic field. In runs A3 and A6, the line widths for the ions are very close to the same when viewed along the mean direction of the magnetic field (Figs. 2 (a) and (b)), but show a noticeable difference when viewed perpendicular to the mean magnetic field (Figs. 2 (d) and (e)). For run A3, the difference between the ion and neutral line widths is about 0.7 $c_s$ and for run A6 it is about 0.5 $c_s$. The spectra shown in Figs. 1 (a) and (b) showed that there is less power in the ions than in the neutrals on length scales shorter than the ambipolar diffusion scale. This results in narrower lines in the ions than in the neutrals, when integrated over all of the length scales. The effect is much more pronounced when looking perpendicular to the mean magnetic field, that is, when the mean magnetic field lies in the plane of the sky. This can be understood intuitively by recalling that the ions are free to move along magnetic field lines, but will feel a force not seen by the neutral molecules when moving normal to the field. The difference between the ion and neutral line widths is not as significant as that seen in Li et al. (2010), however the physical ionization fraction of the clouds they observed is much lower than those of our simulations. There is little to no difference seen between Figs. 2(c) and 2(f), for the lower magnetic field case of run A1. This implies that the difference in the ion and neutral line widths is predominately driven by the presence of a *strong* magnetic field, i.e. it is truly a magnetically-dominated effect. We will elaborate on this in the next sub-section.

**3.4) Linewidth-Size relations**

In this sub-section, we extend the analysis of the previous sub-section to lines on varying length scales. We approximate the emission from each computational zone as an isothermal gas moving with the average velocity of that cell. From this, we produce synthetic line profiles along each individual column of our data cubes. We sum these lines over differently-sized regions of adjacent zones to obtain a combined line profile along each particular line of sight. From each of these we extract the width of the ion and neutral lines. We have thus produced a large sample of line-widths over many length scales spanning a single zone to half of the entire domain.

Fig. 3 shows the collection of these line widths as a function of their length scale, i.e. the linewidth-size relationship. The red squares show the ion line widths and the black diamonds show the neutral line widths. The neutral points are shifted to a slightly higher length scale than the ions for ease of viewing, but they were taken over the same number of zones as the neutrals. No vertical shifts were imparted to any of the data points in Fig. 3. The *x*-axis is scaled to the ambipolar diffusion length scale, thus showing the difference between poorly- and partially-coupled regimes. Figs. 3 (a) – (c) represent the direction along the line of sight of the mean magnetic field, while Figs. 3 (d) – (f) show the direction perpendicular to the mean magnetic field. Figs. 3 (a) and (d) correspond to run A3, Figs. 3 (b) and (e) to run A6 and Figs. 3 (c) and (f) to run A1. The plots in Fig. 3 show a clear difference between the ion and neutral line widths. In particular the lower envelope of the linewidths in the ions is consistently lower than that of the neutrals for all but the longest length scales (which are longer than the driving scale). This is most pronounced in Figs. 3 (d) – (f), where the lines are taken perpendicular to the direction of the mean magnetic field. In particular, compare Figs. 3 (b) and (e), which correspond to run A6. Fig. 3 (b) shows that on all length scales except one, the smallest ion line width is nearly identical to the smallest neutral line width when looking along the direction of the mean magnetic field. However, when looking in the direction perpendicular to the mean magnetic field, i.e. with mean field oriented in the plane of the sky, as is the case for Fig. 3 (e), the smallest ion line width is consistently narrower than the smallest neutral line width.

Our results from Fig. 3 are consistent with the results observed by Li & Houde (2008) in their Fig. 2. In both cases, we see that on every length scale, the ions have a smaller minimum linewidth than the neutrals. This also suggests an additional diagnostic when observing molecular clouds. The difference between the ion and neutral line widths is sensitive to the direction of the mean magnetic field and is most pronounced when observing with the magnetic field in the plane of the sky. The absence of a difference between ion and neutral lines in an observed molecular could indicate that the magnetic field is pointed purely along the line of sight of the observer. Realize that Zeeman splitting observations are only sensitive to the line of sight component of the magnetic field. In contrast, the linewidth-size relationship provides a complementary constraint on the observed magnetic field direction. This suggests that clouds where the linewidth-size relationship shows strong differences between the ions and the neutrals we should expect a weaker signal from Zeeman splitting.

Fig. 4 shows a schematic of the observational cases addressed in this section. In the figure, the Alfvén and magnetosonic waves propagate along the magnetic field, which is drawn along the *x*-axis. These waves cause velocity fluctuations which are transverse to the mean field. As a result, the magnetic field's effect on velocity fluctuations can be picked out most easily along lines of sight that are orthogonal to the mean field, i.e. the *y*- and *z*-directions in the figure. For length scales which are much longer than the ambipolar diffusion scale, the ions and neutrals will be strongly coupled, and little difference should be observed because the fluid will act similarly to single-fluid MHD. For intermediate length scales where the coupling is only partial and on small length scales where the ions and neutrals are not coupled, the difference in velocity will become more pronounced. The structure of the velocity, however, should still be similar when observed along the line of sight of the mean magnetic field.

**4) Probability Distribution Functions of Density**

**4.1) PDF Plots of Density and their Relation to Two-Fluid Turbulence**

Density studies, in particular column densities, are also useful for comparisons between simulation and observation. In this section, we focus on the probability distribution function (PDF) of the ion and the neutral densities. As turbulence progresses, the ions and neutrals can be displaced relative to each other on length scales where they are not strongly coupled, and thus they would present different column densities. As in the previous section, we are interested in the difference between ions and neutrals, and the effect of the magnetic field. Fig. 5 compares the linear PDFs of the ions and neutrals for run A6. The red line represents the ions, and the black represents the neutrals. Each of these PDFs is averaged over 4 different time-points of the simulation, though no significant difference is seen between the different individual PDFs. Fig. 5 (a) shows the PDF when the columns are taken along the direction of the mean magnetic field, and Fig. 5 (b) shows the same when the mean magnetic field is perpendicular to the line of sight. Each of these PDFs is normalized by subtracting the mean value and dividing by the standard deviation. A striking difference is seen in Fig. 5 (a), where the ion PDF has a very different shape than the neutral PDF. The ion PDF is peaked at a lower value and has a longer, shallower tail in the direction of high densities than the neutral PDF. No difference is seen in Fig. 5 (b), when the column densities are taken perpendicular to the mean magnetic field.

This difference is attributable to the fact that the ions are coupled strongly to the magnetic field and are thus are less able to rearrange themselves in either direction, unlike the neutrals. Again, we appeal to Fig. 4 to aid in our explanation. When the Alfvén and magnetosonic waves shown interact non-linearly with one another, they cause more significant density fluctuations in the *yz*-plane. Thus the density distribution in the *yz*-plane will be significantly different from that in the *xy*- and *xz*-planes. Again, on the largest length scales the ions and neutrals will be well coupled and the density differences will be minimal. For scales where the coupling is partial, however, the magnetic field will provide additional support to the ions not seen by the neutrals in the *yz*-plane. Thus, column density distributions observed along a line of sight that is aligned with the field will be different between the ions and the neutrals. This is a result of strong magnetic fields which are capable of resisting the motions of supersonic turbulence.

Fig. 6 shows snapshots of the column densities for run A6. Fig. 6 (a) shows the neutral column density along the mean magnetic field direction. Fig. 6 (b) shows the column densities of the ions in the same direction as (a). Fig. 6 (c) shows the column density of the neutrals perpendicular to the mean magnetic field direction and Fig. 6 (d) shows the column density of the ions perpendicular to the magnetic field. Notice first that in the column density maps taken perpendicular to the mean field, as in Figs. 6 (c) and (d), the basic structure is similar between the ions and neutrals, even if different in detail. However, when the system is observed along the line of sight of the magnetic field, as in Figs. 6 (a) and (b), the neutral column density map is qualitatively different than the ion column density map. In particular, there are many fewer regions of particularly high column density in the ions than in the neutrals. The flux-freezing condition in the ionized fluid prevents higher densities from collecting along columns in the direction of the mean magnetic field, an effect which is not seen in the neutrals. Fig. 6 also shows the effect of the partial coupling of the ion and neutral fluids. The general structure and position of the

higher- and lower-density regions is similar between the ions and the neutrals on the largest scales, but the detailed structure on smaller scales is different. Recall that the ambipolar diffusion length scale is 80 zones for this simulation, or about 0.16 of the entire domain.

The differences in the PDFs are seen in the direction parallel to the mean magnetic field, and no significant differences are noted along the direction perpendicular to the mean field. Therefore, for the remainder of this section, we focus on that direction. Fig. 7 compares the PDFs along the direction of the mean magnetic field from runs A3 and A6. Run A3 is shown in blue and A6 in red. Fig. 7 (a) shows the ion PDF and Fig. 7 (b) shows the neutral PDF. The first thing to note is that the trend shown in Fig. 5 continues, because the PDFs are nearly the same. In Fig. 7 (a), it is seen that the tail of the PDF is higher and shallower for run A6 than for run A3. Fig. 7 (b) shows that the neutral PDFs are nearly identical for both cases. Kowal, Lazarian & Beresnyak (2007) showed that the column density PDF for ideal MHD will have a shallower tail as sonic Mach number increases. The two runs presented here have the same sonic Mach number, but the ions have a slightly shallower tail. This can be understood by recognizing that with increased magnetic field, the ions will have access to faster waves. The neutrals remain unchanged with the changing magnetic field, and thus their PDF seems to be dependent primarily on sonic Mach number. Fig. 8 shows the column density PDFs for run A1 along the direction of the mean magnetic field. Ions are in red and neutrals are in black. In this case, the ions and neutrals have a very similar PDF, suggesting that the effects observed above in runs A3 and A6 only hold when the magnetic field is sufficiently strong. Additionally, the high-end tail of the ion PDF in the direction of the mean field has a much shallower slope for run A1 than for either runs A3 or A6. Since the turbulence is super-Alfvénic in run A1, the mean magnetic field is not strong enough to prevent the supersonic flows from rearranging the ionized fluid.

The effects observed in this section also provide a useful diagnostic in determining the magnetic field orientation. When coupled with another measure of magnetic field, such as Zeeman splitting or dust polarization, the density PDF can shed light on whether the mean magnetic field points along the line of sight or along the plane of the sky. Additionally it is important that the observational direction for the PDF anisotropy is different than the result mentioned in section 3.2 and 3.3 above. As schematically shown in Fig. 4, the density variations will be prominent when observing along the line of sight of the magnetic field, whereas the velocity differences between the ions and neutrals will appear when observing perpendicular to the line of sight of the magnetic field. In summary, the linewidth-size relationship presented in sub-section 3.3 and the density PDF presented in this section complement each other, each providing a dominant observational diagnostic when the other is sub-dominant.

When a molecular cloud is observed exactly along the mean magnetic field, the density PDFs should show prominent differences. In that situation, the linewidth-size relationship for the ions and neutrals will not show a prominent difference. When the molecular cloud is observed orthogonal to the mean magnetic field, the density PDFs should coincide. However, in that situation, the linewidth-size relationship should show prominent differences between the ions and neutrals. We should also point out that this analysis is based on the underlying simplification there there is a single mean magnetic field direction, i.e. the magnetic field is not curved.

**4.2) Moments of the Density PDFs and the Nature of the Two-Fluid Turbulence**

Density and column density PDFs represent one method of determining the sonic Mach number (Kowal et al. 2007) and for testing theories of star formation (Hennebelle & Chabrier 2008). For example, several authors have suggested have suggested that the turbulent sonic Mach number can be estimated from the calculation of the density/column density variance (Padoan et al. 1997; Passot & Vazquez-Semadeni 1998; Price et al. 2011; Burkhart & Lazarian 2012) and the column density skewness/kurtosis (Kowal et al. 2007; Burkhart et al. 2009, 2010). However, all of these studies incorporated only single fluid ideal MHD. As we have shown in the previous section, the PDFs of the neutral and ion column densities will show differences in the tails and peaks of the distribution along sight lines parallel to the mean field. We quantify these effects for column density (represented as $\Sigma$) using a moment analysis investigating the variance ($\sigma^2$), skewness ($\beta$), and kurtosis ($\gamma$) of the distribution and how it depends on the sonic Mach number. Additionally we investigate the variance of the logarithm of the distribution of column densities, i.e. $\zeta = \ln(\Sigma/\Sigma_0)$, which has been shown to depend on the sonic Mach number in isothermal ideal MHD synthetic column density maps as: $\sigma_{2\zeta} = ln(1 + b^2 M_s^2)$, where parameter $b$ depends on the turbulence forcing (see Federrath et al. 2008) and parameter $A$ is a scaling factor between the density and column density distributions (see Burkhart & Lazarian 2012).

We plot the moments vs. sonic Mach number of column density distribution for ions and neutrals for sight lines parallel to the mean field in Fig. 9 and for sight lines perpendicular to the mean field in Fig. 10. We directly calculate the moments (i.e. not from fitting a Gaussian) and average over 4 different snap shots of the turbulence. From top panel to bottom they are the variance of the logarithm of the column density distribution normalized by its mean value ($\zeta = \ln(\Sigma/\Sigma_0)$, the variance of the linear distribution normalized by its mean value, the skewness ($\beta$) of the linear distribution, and finally the kurtosis of the linear distribution ($\gamma$). In all cases we include ideal isothermal MHD simulations with resolution $512^3$ from Burkhart & Lazarian 2012 (BL12) for reference. For the variance tracers, we include the reported line of best fit from BL12 for ideal MHD simulations as a dashed black line.

For the case of parallel sight lines in Fig. 9, the skewness ($\beta$) and kurtosis ($\gamma$) show what we visually expect from the previous sub-section. Namely, that the ions (shown in red) have skewness (tails of the PDF) and kurtosis (peak of the PDF) consistently higher than the neutrals within the error bars. Interestingly, the variance, both of the linear and logarithm of the distribution, is nearly the same in the ions and neutrals for the simulations with sub-Alfvénic turbulence (A3 and A6). A1 shows variance differences, with ions having wider PDFs then neutrals. This is due to the fact that the Alfvén and slow waves are significantly damped, leaving the compressive fast mode to create a bulk increase in the moments of the ions, including the variance. Furthermore, for super-Alfvénic turbulence where the influence of the magnetic field is diminished, the ion motion is not as restricted as in the case of the high field situation. The neutral variance also increases, but this is due to the higher sonic Mach number for this model.

For sight lines perpendicular to the mean magnetic field (shown in Fig. 10), we again see what is visually expected for skewness and kurtosis from the previous sub-section. Although differences exist between ion and neutral skewness/kurtosis, they are generally not significant within the error bars. Furthermore, no clear trend is seen between ions and neutrals. For example, model A6 shows higher neutral skewness and kurtosis than in the ions. This analysis again shows that only sight lines parallel to the mean field will show significant enhancement of ion skewness and kurotis for sub-Alfvénic turbulence. Interestingly, the variance shows clear differences between ions and neutrals well within the error bars with ions having consistently higher variance. This is in contrast to sight-lines parallel to the mean field, where the variance of the ions/neutrals was nearly the same for the A3 and A6 models. We note that for A1 (super-Alfvénic), the ions always show moments that are significantly higher than the neutrals, regardless of sight line. This is due to the fact that the fast mode is the only MHD mode propagating on length scales shorter than the ambipolar diffusion length scale. The longer propagation of this compressible mode increases the moments in all cases.

## 5) Discussion and Conclusions

We have presented high resolution simulations of two-fluid magnetohydrodynamic (MHD) turbulence at resolutions as large as $512^3$ zones. The simulations are supersonic and range from mildly sub-Alfvénic to super-Alfvénic. These parameter choices are in keeping with the conditions that arise in molecular clouds. Such turbulence is observed on large scales in molecular clouds, all of which are partially ionized owing to their low temperature and relatively high density. On length scales that are larger than a parsec, the ions and neutrals are very strongly coupled and the two fluid turbulence masquerades as a single fluid MHD turbulence. Indeed, our simulations make this clear. On smaller scales, approaching $L_{AD}$, the turbulence changes character owing to the fact that the Alfvén and fast magnetosonic wave families change their propagation characterstics and are damped. For fiducial cloud parameters, $L_{AD} \sim 0.05$ pc. Analytical theory (Balsara 1996) finds that on scales ranging from $10\, L_{AD}$ to $L_{AD}$, the MHD waves in the ions undergo substantial damping. These waves have a preferential anisotropic propagation along the magnetic field lines. We find that the turbulence in the ions does propagate to length scales somewhat below $L_{AD}$, till it is fully damped. This is consistent with the analysis in Balsara (1996) which shows that there is a Reynolds number-like dimensionless number and the turbulence in the ions propagates down to those scales and dies out only when $\text{Re}_{AD} \sim 1$. For length scales smaller than $L_{AD}$, the turbulence can still have strong velocity fluctuations, even though the waves become non-propagating. In that sense, it can still be a substantial source of reconnection in molecular cloud cores (Lazarian 2011) since the magnetic field is still enslaved to the ions on those smaller length scales. Thus our simulations reconcile the turbulence-only scenarios with the alternative scenarios for star formation (which insist on magnetic field removal from cores), forming as it were a continuous bridge from one model to the other.

We show that several diagnostics can be extracted from our simulations, confirming recent results on linewidth-size relationsips observed in optically thin isophotologous lines (Li & Houde 2008, Li *et al.* 2010, Hazareh *et al.* 2012). The linewidth-size relations can be interpreted in terms of damping of magnetosonic and Alfvén waves in the turbulence (Balsara 1996). Once these families of propagating waves is denied to the turbulence in the ions, that component of the turbulence rapidly dies out. It is important to realize though that the turbulence dies out over a large range of length scales spanning $\sim 10\, L_{AD}$ to a small fraction of $L_{AD}$. The turbulence in the neutrals is unimpeded on these length scales. Over this range of length scales, the ions and neutrals become weakly coupled and the waves that sustain the turbulence propagate mostly as MHD waves in the ions and hydrodynamic sound waves in the neutrals. As a result, over this substantial range of lengths, the velocities in the ions and neutrals differ considerably. It is because of this considerable difference over a large range of length scales that the linewidth-size relationship has been verified over a substantial range of length scales.

We focus on waves with length scales spanning $\sim 10\, L_{AD}$ to a small fraction of $L_{AD}$. It is important to realize that the character of wave propagation in the ions changes considerably, with the Alfvén and magnetosonic waves propagating at speeds that are comparable to the *Alfvén speeds in the ion component*. Any fast reconnection that takes place with these waves should be unusually efficient. The polarization properties of these waves is also very different from that of sound waves. These waves should be propagating preferentially in the direction of the magnetic field with velocity fluctuations that are orthogonal to the direction of the mean magnetic field. As a result, velocity differences between ion and neutral velocities should show up most prominently along lines of sight that are orthogonal to the magnetic field. The transverse velocity fluctuations should produce density fluctuations that show up most prominently when viewed along the field lines. Based on these insights, we offer up the following observational diagnostics:

**1)** We find that the linewidth-size relationship should show a prominent difference between ions and neutrals when the line of sight is orthogonal to the mean field.

**2)** We also find that the density PDFs and their derived diagnostics should show prominent differences between the ions and neutrals when the line of sight is parallel to the mean field. The skewness and kurtosis of the linear PDFs of the ion densities was significantly higher than that of the neutrals when observed along the line of sight of the mean magnetic field. A significant difference was not seen perpendicular to the mean field direction.

**3)** When the magnetic field makes an angle to the line of sight, both linewidth-size differences and density PDF differences should be visible in the ions and neutrals. If the magnetic field curves inside the cloud, one of those diagnostics could become prominent in one location of the cloud with the other diagnostic becoming prominent in another location.

The diagnostics presented here should be easy for observers to test. The present analysis is predicated on observations in optically thin lines. It is also predicated on there being a uniform mean magnetic field that permeates the cloud.

**Acknowledgements**

DB acknowledges support via NSF grant NSF-AST-1009091. DB also acknowledges support via NASA grants from the Fermi program as well as NASA-NNX 12A088G.

DB gratefully acknowledges the use of NSF-funded XSEDE resources on NICS Kraken where a majority of the simulations reported in this paper were performed. The use of computer resources at Notre Dame's Center for Research Computing is also gratefully acknowledged.

BB acknowledges support from the NASA Wisconsin Space Grant Institution. AL and BB thank both NSF AST 0808118 and the the center for Magnetic Self-Organization in Astrophysical and Laboratory Plasmas for financial Support.

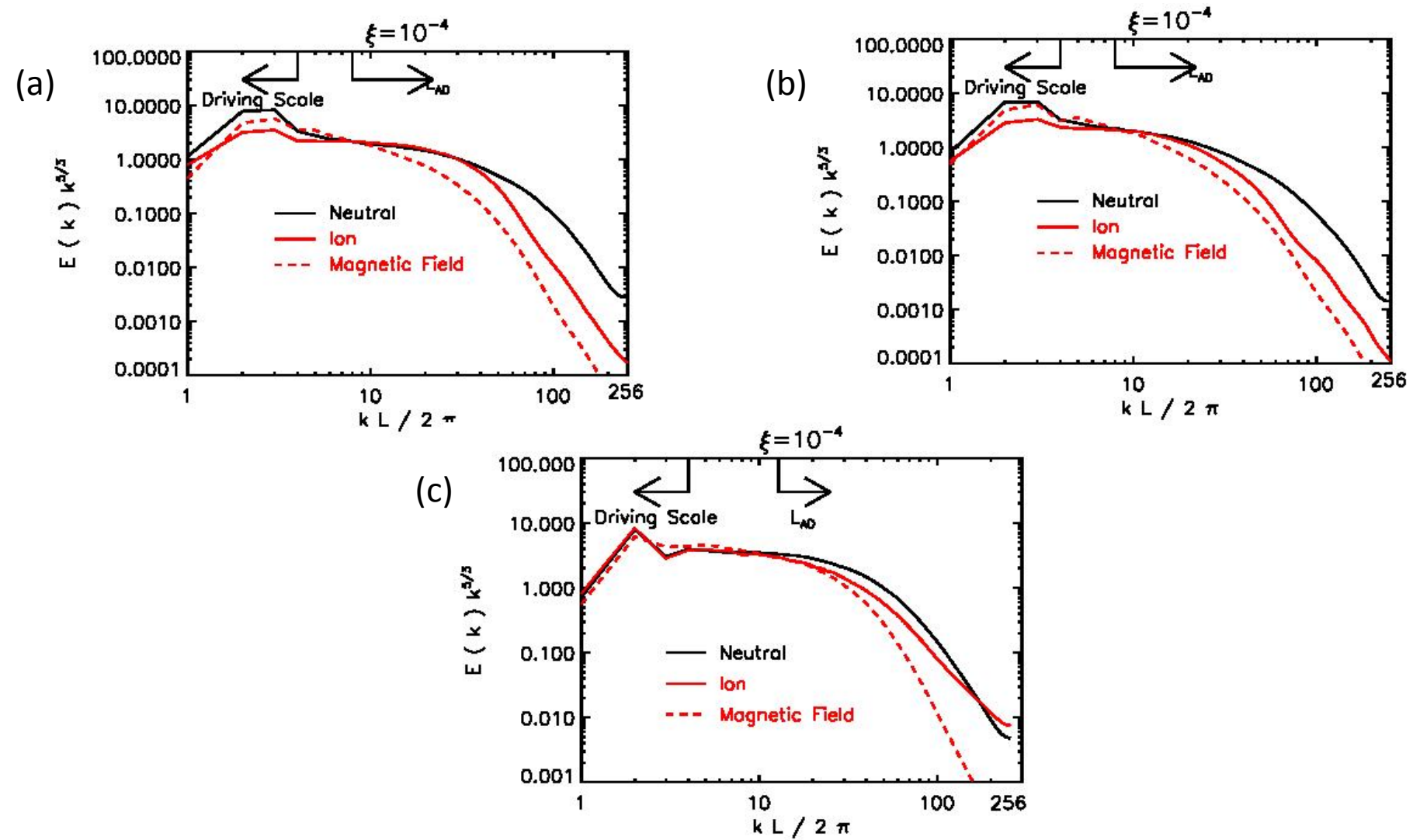

Figure 1. The 1D compensated energy spectrum of the three simulations. Figure 1(a) shows run A3, (b) shows run A6 and (c) shows run A1. The black line shows the neutral kinetic energy, the red line shows the ion kinetic energy and the red dashed line shows the magnetic energy. The spectra are all multiplied by a factor of $k^{5/3}$. The driving scales and ambipolar diffusion scale is shown on each plot.

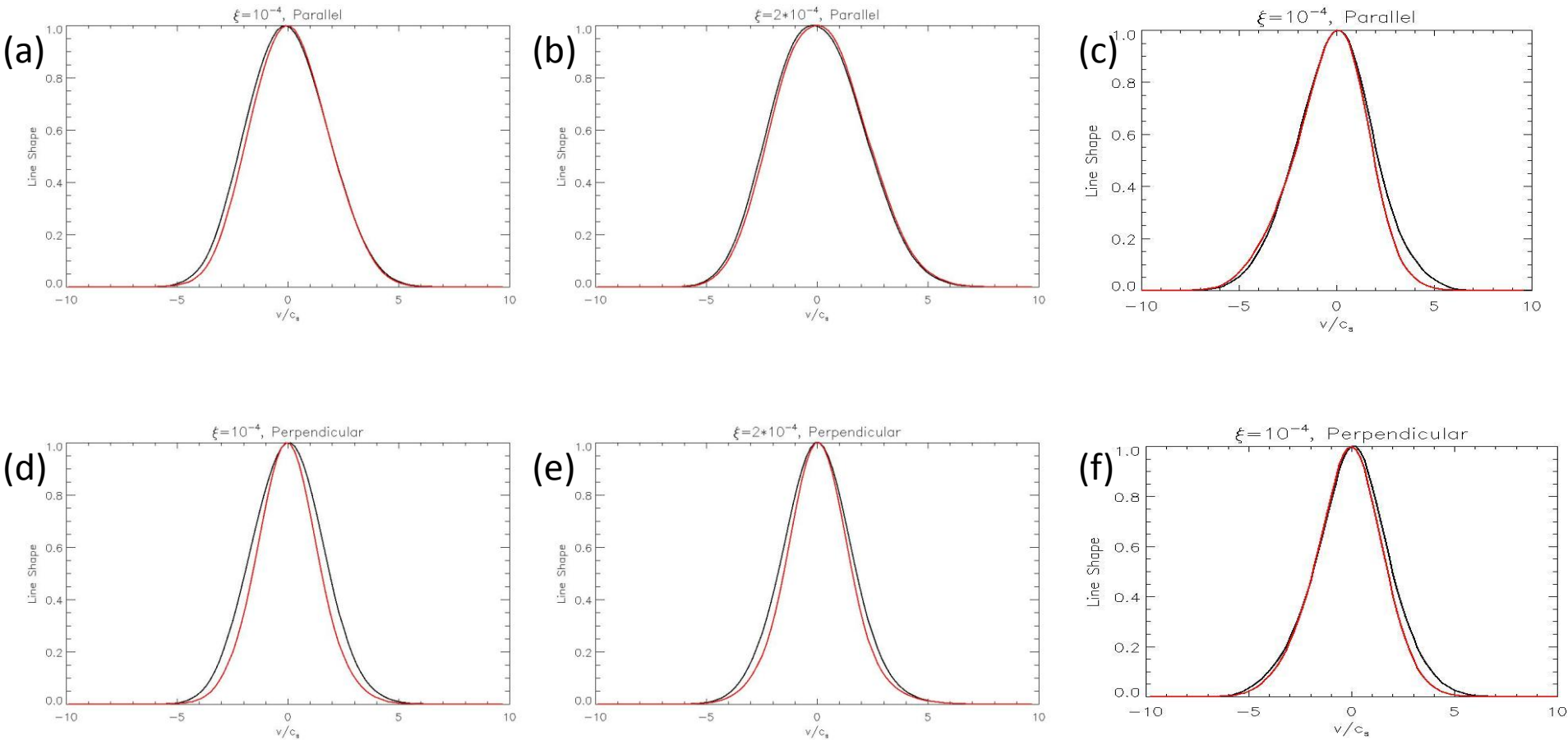

Figure 2. Synthetic line widths for our simulations taken over the entire computational domain. Plots (a) – (c) show the profiles along the direction of the mean magnetic field, and plots (d) – (f) show the profiles perpendicular to the mean magnetic field. Plots (a) and (d) correspond to run A3, plots (b) and (e) to run A6 and plots (c) and (f) to run A1. The black line shows the profile for the neutrals and the red the profile for the ions.

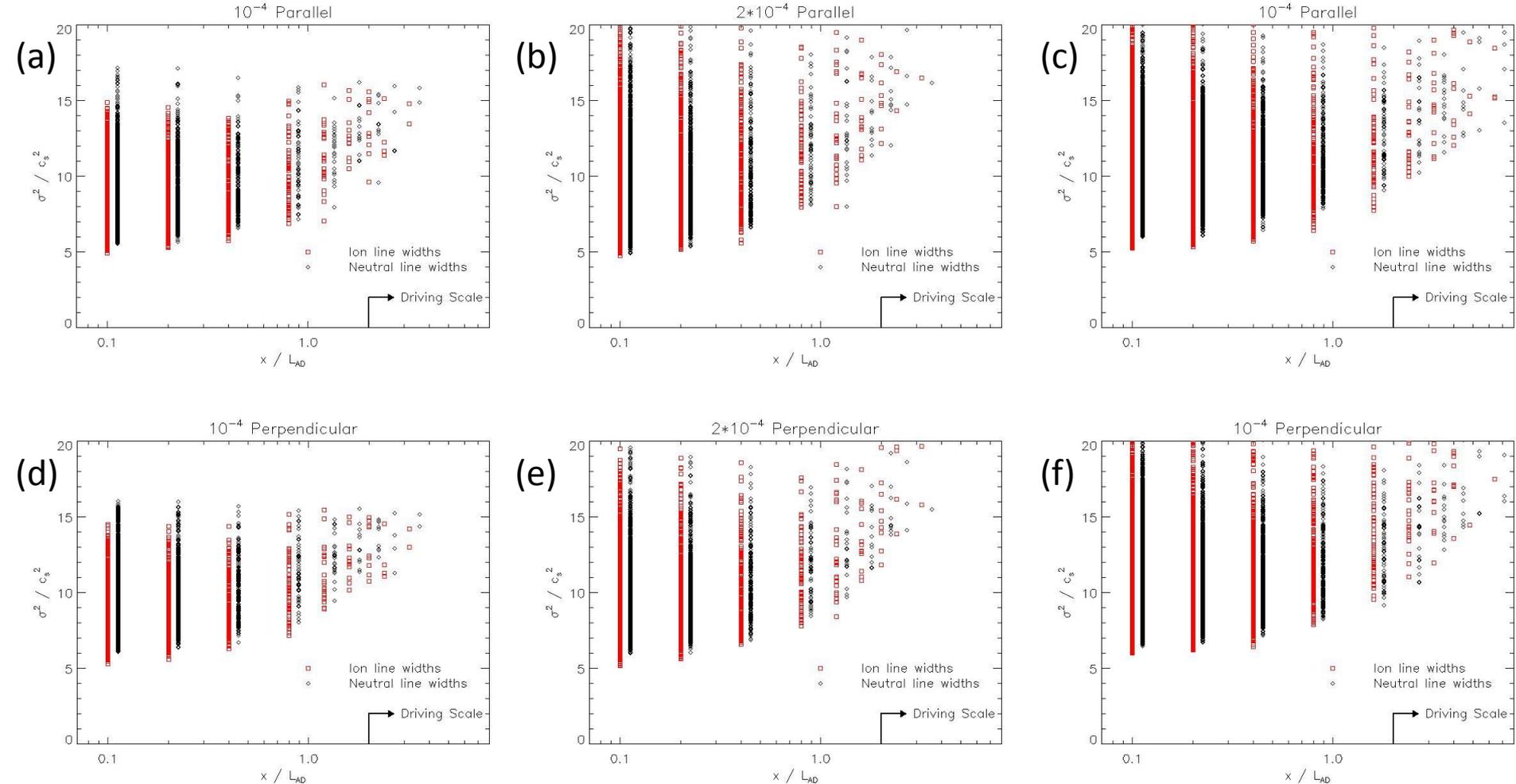

Figure 3. The linewidth-size relationship for our simulations. The ion line widths are shown as red squares, and the neutrals as black diamonds. The neutrals are shifted slightly from the ions so they can be clearly seen, but they both represent the same length scales. Plots (a) – (c) represent the direction along the line of sight of the mean magnetic field, and (d) – (f) the direction perpendicular to the mean magnetic field. Plots (a) and (d) correspond to run A3, plots (b) and (e) to run A6 and plots (c) and (f) to run A1. The x-axis is scaled to the ambipolar diffusion scale, and the driving scale is labeled in each plot.

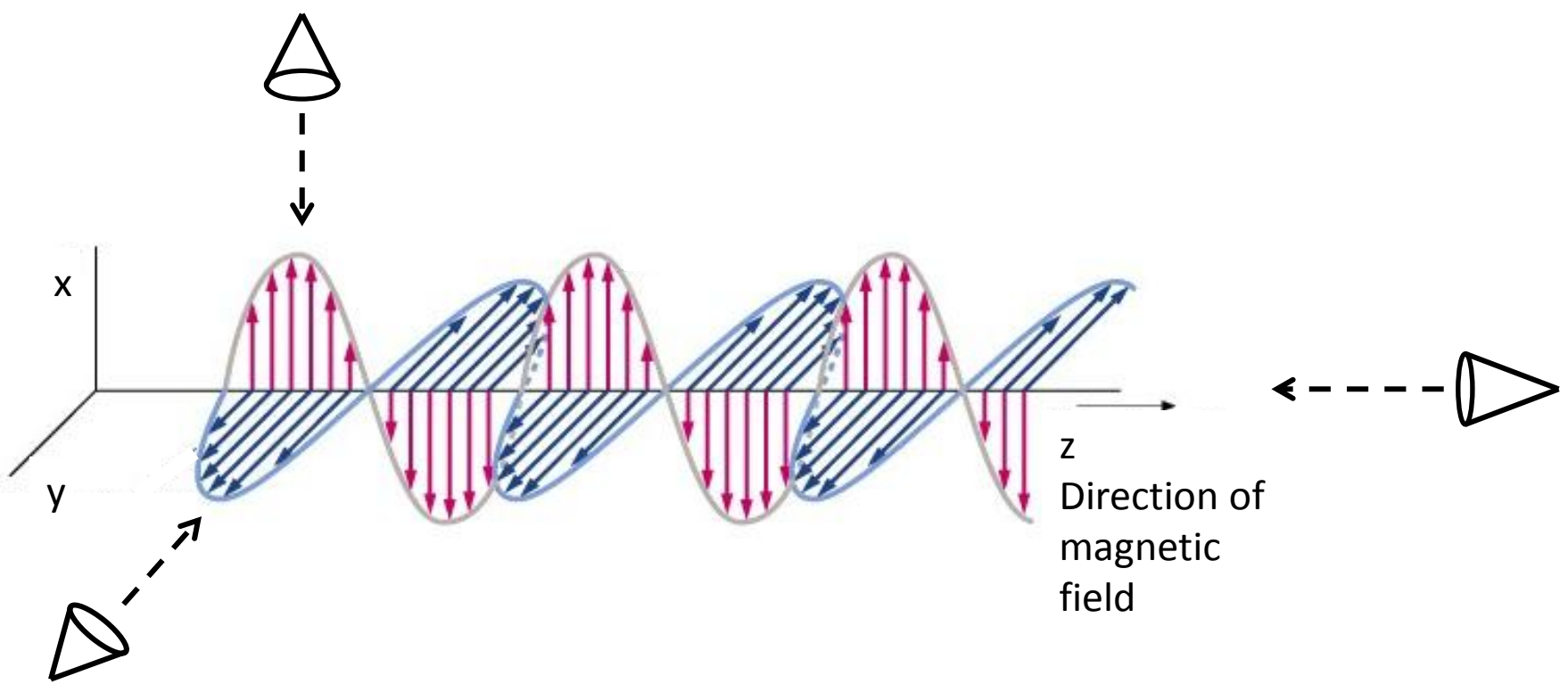

Figure 4. The presence of a strong magnetic field alters MHD wave propagation. The Alfven and magnetosonic waves propagating along the field are shown here. They cause velocity fluctuations transverse to the mean field. As a result, the magnetic field's effect on velocity fluctuations is picked out most easily along lines of sight that are orthogonal to the mean field, i.e. the x- and y-directions in the figure. However, when these waves interact non-linearly with one another, they cause a preponderance of density fluctuations in the xy-plane. Thus the density distribution in the xy-plane will be significantly different from that in the xz- and yz-planes. Consequently, density distributions observed along a line of sight that is aligned with the field (i.e. in the z-direction) will be significantly different from the density distribution along the other two lines of sight. On length scales where the ions and neutrals are weakly coupled, no such anisotropy in velocity or density is expected for the neutral fluid.

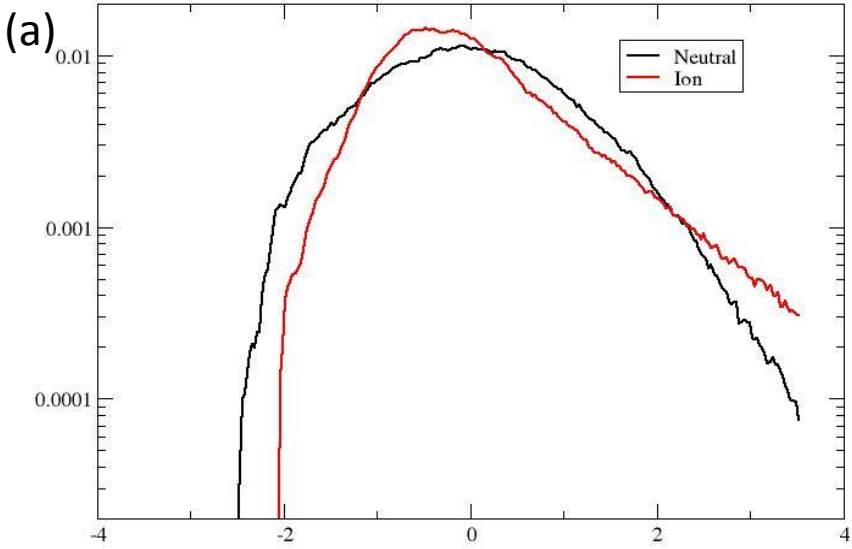 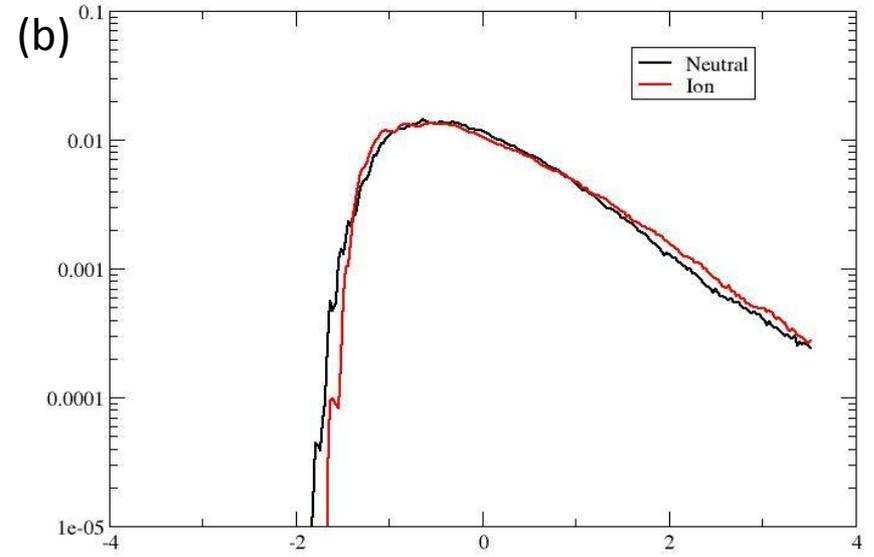

Figure 5. The column density PDFs of run A6. Plot (a) shows the PDF along the direction of the mean magnetic field, and (b) shows the same perpendicular to the mean field. Ions are shown in red and neutrals in black.

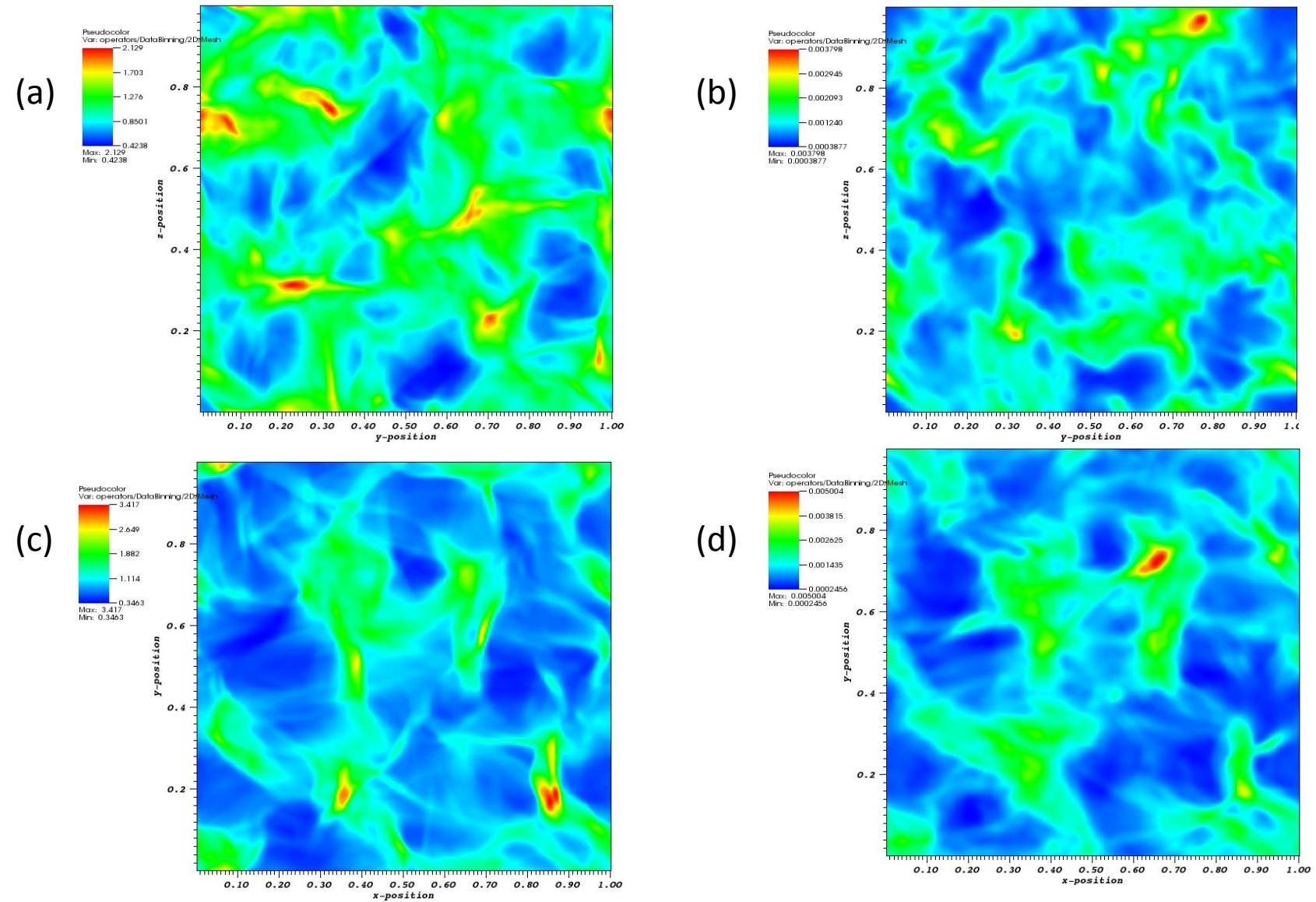

Figure 6. Column density maps for run A6. Plot (a) shows the neutral density along the direction of the mean magnetic field, (b) shows the ions along the same line of sight, (c) shows the neutral column density perpendicular to the mean field, and (d) shows the ion column density along the same line of sight as (c).

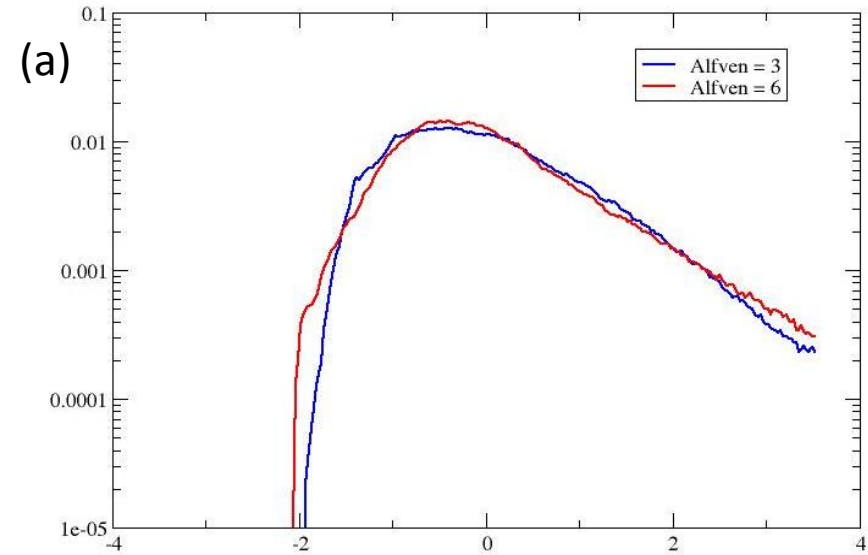 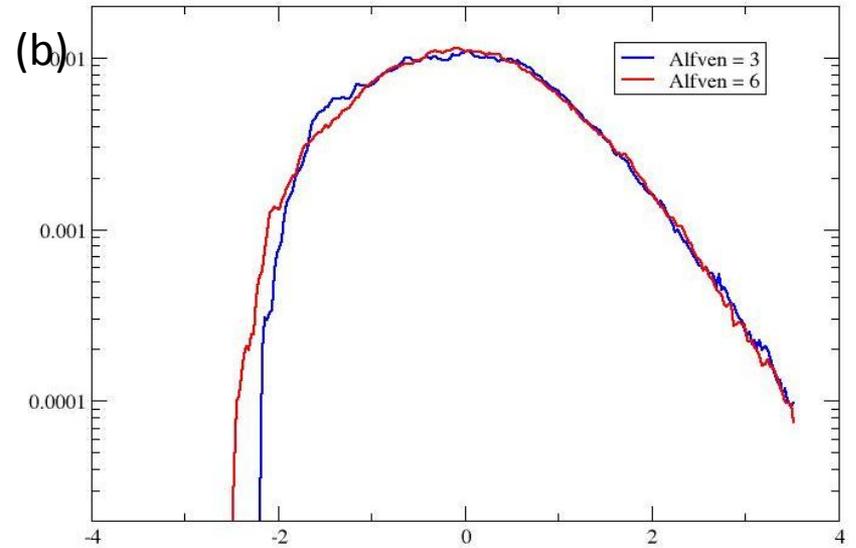

Figure 7. A comparison of the column density PDFs for runs A3 and A6 along the direction of the mean magnetic field. Plot (a) shows the PDF of the ions, and (b) shows the same for the neutrals. Run A3 is in blue and A6 in red.

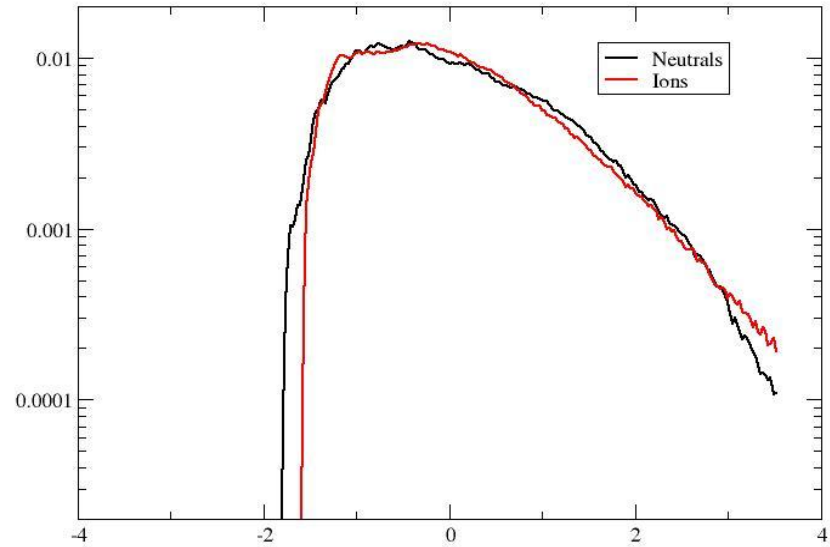

Figure 8. The column density PDFs for run A1 along the direction of the mean magnetic field. Ions are in red and neutrals are in black.

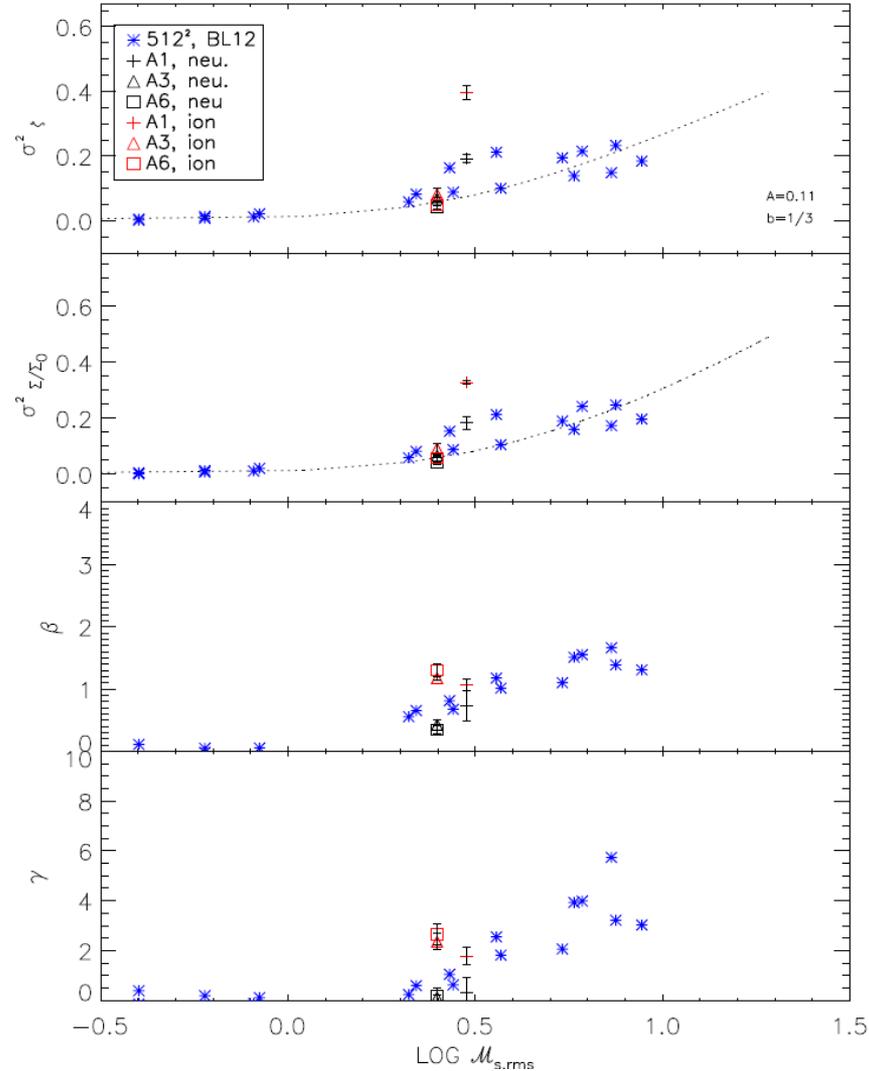

Figure 9. The scalar moments of the linear PDFs taken along the direction parallel to the mean magnetic field. The blue stars and the dotted line correspond to the ideal MHD simulations and best fit presented in Burkhart & Lazarian (2012). For the simulations presented in this paper, the plus signs represent run A1, the triangles run A3 and the squares run A6, with the red points representing the ions and the black points representing the neutrals. The top panel shows the variance of the logarithm of the column density, the second panel shows the variance of the column density, the third panel shows the skewness and the bottom panel shows the kurtosis.

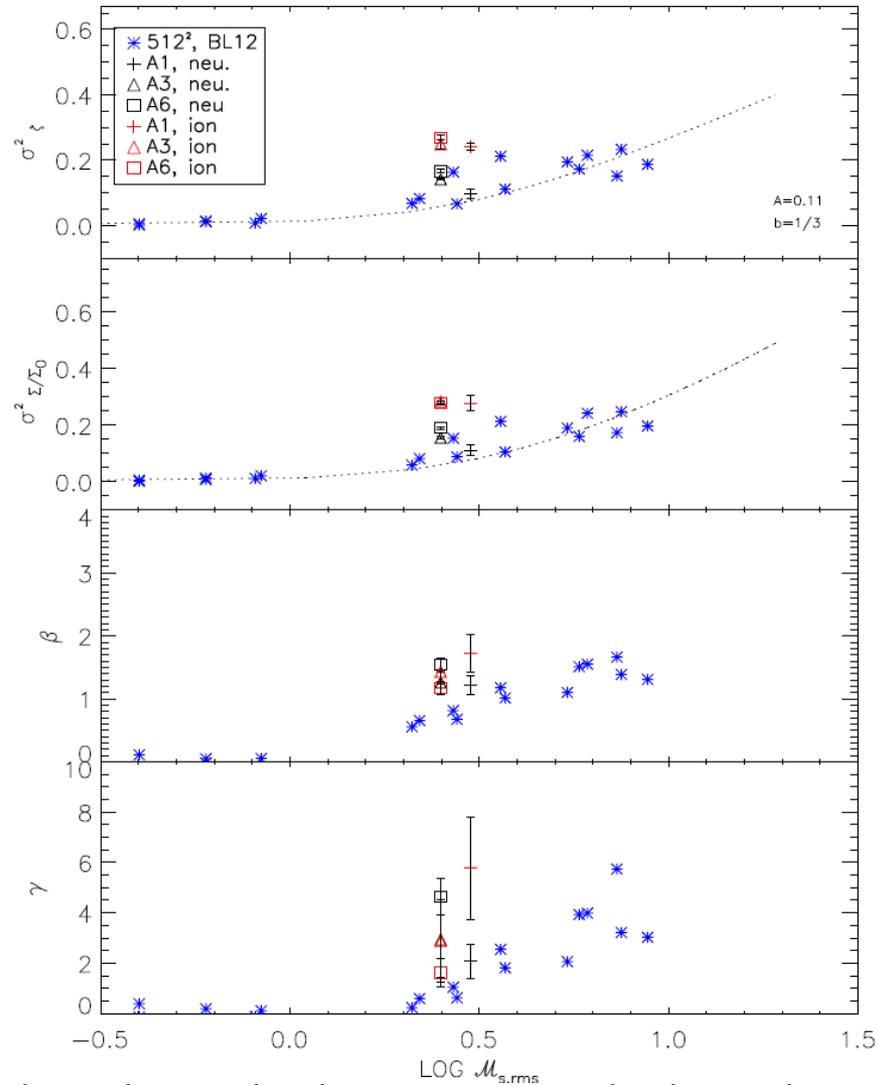

Figure 10. The same as Fig. 9, but taken in the direction perpendicular to the mean magnetic field. The blue stars and the dotted line correspond to the ideal MHD simulations and best fit presented in Burkhart & Lazarian (2012). For the simulations presented in this paper, the plus signs represent run A1, the triangles run A3 and the squares run A6, with the red points representing the ions and the black points representing the neutrals. The top panel shows the variance of the logarithm of the column density, the second panel shows the variance of the column density, the third panel shows the skewness and the bottom panel shows the kurtosis.